\documentclass[prl,aps,twocolumn,showpacs,superscriptaddress,floatfix]{revtex4}
\usepackage{graphicx}
\usepackage{wasysym}

\newcommand{\beq}{\begin{equation}}
\newcommand{\eeq}{\end{equation}}
\newcommand{\bea}{\begin{eqnarray}}
\newcommand{\eea}{\end{eqnarray}}
\newcommand{\cI}{{\cal I}}
\newcommand{\cS}{{\cal SF}}

\begin{document}

\title{Spin liquid phase in a $S=1/2$ quantum magnet
on the kagome lattice}

\author{S. V. Isakov}
\affiliation{Department of Physics, University of Toronto, Toronto,
Ontario M5S 1A7, Canada}
\author{Yong Baek Kim}
\affiliation{Department of Physics, University of Toronto, Toronto,
Ontario M5S 1A7, Canada}
\author{A. Paramekanti}
\affiliation{Department of Physics, University of Toronto, Toronto,
Ontario M5S 1A7, Canada}

\date{\today}

\begin{abstract}

We study a model of hard-core bosons with short-range repulsive interactions
at half-filling on the kagome lattice. Using quantum Monte Carlo (QMC)
numerics, we find that this model shows a continuous superfluid-insulator
quantum phase transition (QPT), with exponents $z=1$ and $\nu\approx 0.67(5)$.
The insulator, $\cI^*$, exhibits short-ranged density and bond correlations,
topological order and exponentially decaying spatial vison correlations, all
of which point to a $Z_2$ fractionalized phase. We estimate the vison gap in
$\cI^*$ from the temperature dependence of the energy. Our results, together
with the equivalence between hard-core bosons and $S=1/2$ spins, provide
compelling evidence for a spin-liquid phase in an easy-axis spin-$1/2$ model
with no special conservation laws.
\end{abstract}

\pacs{75.10.Jm, 05.30.Jp, 71.27.+a, 75.40.Mg}

\maketitle

Spin liquid states, or strongly correlated quantum paramagnets that 
preserve all lattice symmetries, were proposed long ago as plausible 
candidates for the ground state of frustrated quantum magnets 
\cite{fazekas74}. In recent years, several frustrated antiferromagnets
have been discovered which appear to either have a spin-liquid 
ground state (such as the $S=1$ kagome lattice system Nd$_3$Ga$_5$SiO$_{14}$ 
\cite{spinliq:ndlangasite} or the $S=1/2$ triangular lattice Mott 
insulator $\kappa$-(ET)$_2$Cu$_2$(CN)$_3$ \cite{spinliq:organic})
or are proximate to a spin-liquid phase (such as the $S=1/2$
stacked triangular magnet Cs$_2$CuCl$_4$ \cite{spinliq:cscucl}), 
thus lending support to this paradigm.
On the theoretical front, there has been considerable progress in
understanding the effective field theories and properties of such spin-liquid
phases \cite{wen,subir,senthil,motrunich:senthil}, showing 
that the excitations in this phase carry fractional quantum numbers and 
interact with emergent gauge fields.
However, there is the pressing issue that most microscopic models
which can be shown to have a spin-liquid phase have either very unusual
interactions \cite{kitaev:wen}
or local Hilbert space constraints. An example of the
latter are particular triangular \cite{moessner} and kagome lattice 
\cite{misguich} quantum dimer models which have
dimer-liquid ground states but are not directly related to any microscopic 
spin Hamiltonians.

In a significant development, Balents {\it et al} \cite{balents1}
proposed a model of $S=1/2$ spins on the kagome lattice
\beq
  H_{\text{XXZ}}=
    - J_{\perp} \sum_{\hexagon} [(S^x_{\hexagon})^2+(S^y_{\hexagon})^2-3]
    +J_z \sum_{\hexagon} (S^z_{\hexagon})^2,
\label{eq:xxzmod}
\eeq
where $S^a_{\hexagon}=\sum_{i\in\hexagon}S^a_i$ is a sum over the
six spins on each hexagon of the kagome lattice unit cell,
$\sum_{\hexagon}$ denotes a sum over all hexagons on the lattice.
This model is easily seen to be a
short-range anisotropic XXZ model, with only the first, second and third 
neighbor interactions being nonzero and equal to each other.
Analyzing this model \cite{balents1} for $J_\perp<0$ and 
$J_z/|J_\perp| \gg 1$, they showed that it directly {\it maps} onto an 
effective triangular lattice quantum dimer model with three dimers touching 
each site. In spin language, this effective model takes the form of a
ring-exchange model, with an exchange scale $J_{\rm ring}=J^2_\perp/J_z$,
which describes quantum dynamics in the Hilbert space with 
$S^a_{\hexagon}=0$ on each hexagon, the local
constraint arising from energetic considerations at large $J_z/J_\perp$. 
Supplementing this
model with an additional four-site (Rokhsar-Kivelson (RK) \cite{rk})
potential term of strength $v_{4}$ was shown to lead to a
spin-liquid for $v_4=J_{\rm ring}$, which was argued to be stable for small 
deviations $v_4<J_{\rm ring}$. Later exact diagonalization (ED) numerics 
\cite{balents2} showed that the ring-exchange model appears to be in this 
spin-liquid phase down to $v_4=0$, but only system sizes upto 20 unit cells
could be explored.
A numerical study of a related rotor model on a decorated
square lattice \cite{senthil:motrunich} also led to a $Z_2$ spin liquid, but
the corresponding $S=1/2$ model was not studied.

In this paper, we show the presence of a spin-liquid phase in a $S=1/2$
model on the kagome lattice, by studying numerically the Hamiltonian
(\ref{eq:xxzmod}) with $J_\perp > 0,  J_z > 0$. Since the ring-exchange model
is independent of the sign of $J_\perp$, we expect to recover the 
physics of the ring-exchange model at large $J_z$, but without imposing
Hilbert space restrictions by hand. On the technical side, the choice of
$J_\perp>0$ eliminates the QMC sign problem. This permits us to study
very large systems at very low temperature using a generalized 
stochastic series expansion QMC algorithm \cite{sse,footnote.sseplaq}
and, thus, to go significantly beyond earlier work on this class of models.

By the mapping between $S=1/2$ spins and hard-core bosons, model 
(\ref{eq:xxzmod}) is equivalent to a hard-core boson model with short-range
repulsion at half-filling
\beq
  H_{\text{b}}=
    -t \sum_{(i,j)}
      \left( b^{\dag}_i b_j + \text{H.c.} \right)
    +V \sum_{\hexagon} (n_{\hexagon})^2
    - \mu \sum_{i} n_i,
\label{eq:bosonmod}
\eeq
where $b^{\dag}_i$($b_j$) is the boson creation(annihilation) operator,
$t = J_\perp > 0$ is the hopping amplitude, $V=J_z > 0$ is the
repulsion strength, $n_i=b^{\dag}_i b_i$ is the number operator, and
$\mu=12 J_z$ is the chemical potential. The hopping term connects only the
first, second and third neighbors. We begin by showing that model
(\ref{eq:bosonmod}) exhibits a superfluid-insulator transition at
$(V/t)_c \approx 19.8$. We then turn to the nature of the insulating phase
and show that it is a topologically ordered $Z_2$ Mott insulator. 
We thus arrive at the result that for $J_\perp>0$, model 
(\ref{eq:xxzmod}) exhibits a ferromagnetic phase at small $J_z$ and 
a $Z_2$ fractionalized spin-liquid at large $J_z$, separated by a
continuous QPT at $(J_z/J_\perp)_c \approx 19.8$.

\begin{figure}[t]
\includegraphics[width=3.0in]{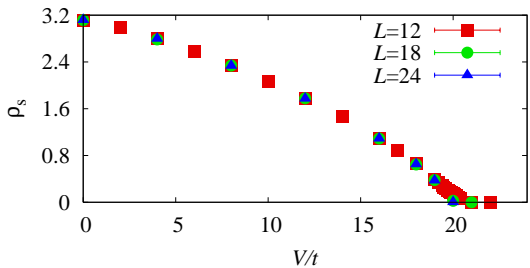}
\caption{ (color online). The superfluid density $\rho_s$ versus
$V/t$, at $T=t/9$ for different linear system sizes $L$. 
}
\label{fig:rhos}
\vskip -0.2cm
\end{figure}

{\it Superfluid-insulator transition}.---For small values of $V/t$, we expect
the ground state of (\ref{eq:bosonmod}) to be a superfluid ($\cS$). We 
confirm this
by measuring the superfluid density $\rho_s$ through winding number
fluctuations  $W_{a_{1,2}}$~\cite{windingnumber} in each of the two lattice
directions, with $
\rho_s=(1/2\beta t) (\langle W_{a_1}^2\rangle + \langle W_{a_2}^2 \rangle)$,
where $\beta$ is the inverse temperature. As seen from Fig.~\ref{fig:rhos}, 
for small $V/t$, $\rho_s$ is large (its value agrees with
mean field estimates \cite{unpub}). $\rho_s$ decreases with
increasing $V/t$, eventually vanishing for $V/t\gtrsim 20$ suggesting a phase
transition to an insulating phase ($\cI^*$). This behavior is
in sharp contrast to a nearly identical model where the hopping and repulsive
interactions are restricted to the nearest neighbor only --- in that case a
uniform superfluid persists for arbitrarily large $V/t$ \cite{kagome:nn}. 
The absence of a jump in $\rho_s$ in Fig.~\ref{fig:rhos}  suggests that the
$\cS-\cI^*$ transition is continuous.

\begin{figure}[t]
\includegraphics[width=3.0in]{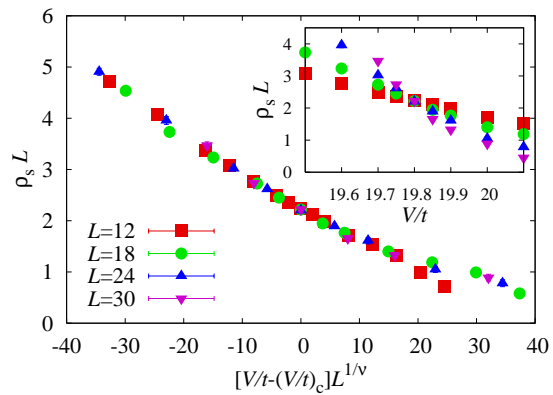}
\caption{ (color online).
Data collapse of the superfluid density $\rho_s$ for $\beta/L=4/(3t)$,
$(V/t)_c=19.80(2)$, and $\nu=0.67(5)$. Inset: Finite size scaling of 
$\rho_s$ for $\beta/L=4/(3t)$.
}
\label{fig:rhos:collapse:scaling}
\vskip -0.2cm
\end{figure}

In the vicinity of a continuous QPT, we expect
$
  \rho_s=L^{-z} F_{\rho_s}(L^{1/\nu}(K_c-K), \beta/L^z),
$
where $F_{\rho_s}$ is the scaling function, $L$ is the linear system size,
$z$ the dynamical critical exponent, $\nu$ the correlation length exponent,
and $(K_c-K) \propto (V_c-V) /t$ is the distance to the critical point.
Thus if we
plot $\rho_s L^z$ as a function of $V/t$ at fixed aspect ratio $\beta/L^z$,
the curves for different system sizes should intersect at the critical point.
The inset of Fig.~\ref{fig:rhos:collapse:scaling} shows such a plot for
$z=1$ and $\beta/L=3/(4t)$ with a clear crossing point
at $(V/t)_c \approx 19.8$.
To obtain the correlation length exponent $\nu$, we plot 
$\rho_s L$ as a
function of $[(V/t)_c-V/t]L^{1/\nu}$ for different system sizes.
It follows from the above scaling relation
that the curves for different system sizes
should collapse onto a universal curve $F_{\rho_s}$ for a properly chosen
$(V/t)_c$ and $\nu$. In
Fig.~\ref{fig:rhos:collapse:scaling}, we show such a data collapse for
$(V/t)_c=19.80(2)$ and $\nu=0.67(5)$. The error bars are estimated from the
stability of the collapse towards varying the parameters. 
To summarize, we find a continuous $\cS-\cI^*$ transition with exponents
$z=1$ and $\nu=0.67(5)$. We next examine the nature of the insulator $\cI^*$.

\begin{figure}[t]
\includegraphics[width=1.5in]{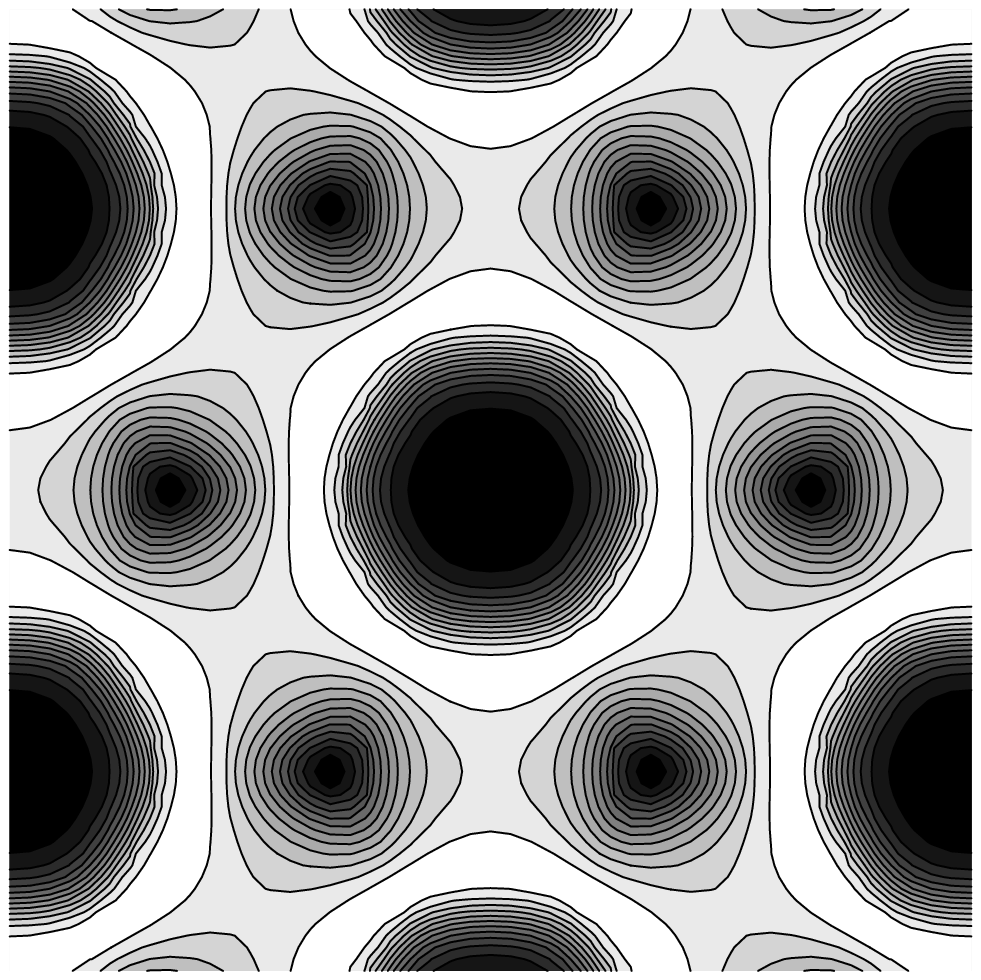}
\includegraphics[width=1.5in]{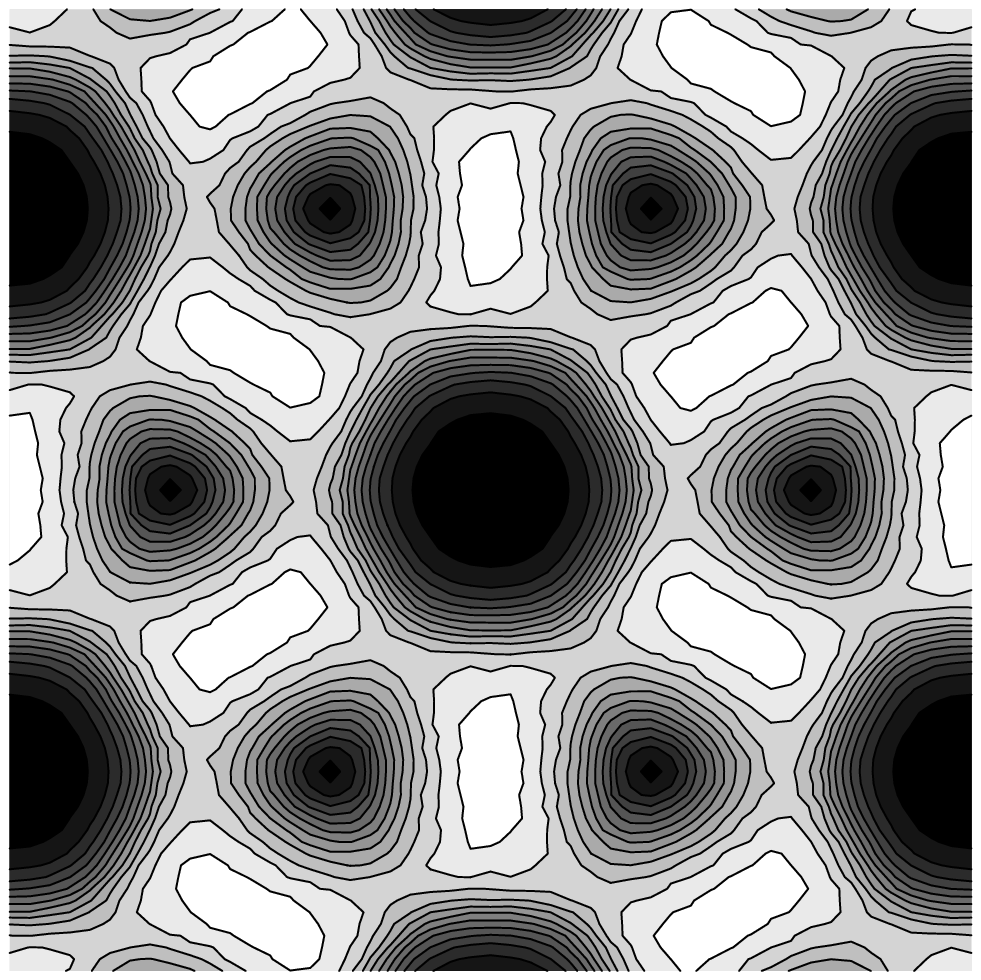}
\caption{ Contour plots of the equal-time density structure factor ($L=24$)
for the classical model (with $t$=0) \cite{3dimer} (left panel) at $T=0$
and in the quantum insulator (right panel) for $V/t=20, T=t/12$. Axes range
from  $-2\pi$ to $2\pi$.
}
\label{fig:corrq}
\vskip -0.2cm
\end{figure}

{\it $\cI^*$ does not break lattice symmetries}.---For a system of bosons,
flux-threading arguments \cite{oshikawa, ashvin} show
that an insulating state at half-filling could
either be a conventional state with broken lattice symmetries or must
{\em necessarily} have topological order (which means the
ground state degeneracy depends on the topology of the system). We rule
out the first possibility here by studying correlation functions in the
insulating state.
We look for signatures of diagonal (density) ordering by studying the 
equal-time density structure factor
$S({\mathbf q})/N=\langle\rho^{\dagger}_{{\mathbf q}\tau}
    \rho_{{\mathbf q}\tau} \rangle,
$
where
$\rho_{{\mathbf q}\tau}=(1/N)\sum_i\rho_{i\tau}\exp(i{\mathbf q}{\mathbf r_i})$
and $\rho_{i\tau}$ is the boson density at site $i$ and imaginary time
$\tau$. Fig.~\ref{fig:corrq} compares a contour plot of the equal time 
structure factor to that of the classical model (with $J_\perp=0$) which is
known to have short-range correlations \cite{3dimer}. The striking similarity
between the two suggests short-ranged density correlations in the full quantum
ground state of the insulator. We confirm this via a careful examination of
the equal-time structure factor
on large system sizes. Upto $L=48$ we do not find any Bragg peaks which rules
out the possibility of density ordering even with moderately large unit
cells. We also examine for possible bond ordering in $\cI^*$ by  computing the
bond-bond structure factor
$
S_{\text b}({\mathbf q})/N=\langle B^{\dagger}_{{\mathbf q}\tau}
B_{{\mathbf q}\tau} \rangle,
$
where $B_{{\mathbf q}\tau}=(1/N)\sum_\alpha B_{\alpha \tau}
\exp(i{\mathbf q}{\mathbf r_\alpha})$ is summed over the bond index $\alpha$
connecting spins $i$ and $j$, and
$B_{\alpha(i,j), \tau}=t(b^{\dag}_i b_j + b_i b^{\dag}_j)$ is the off-diagonal
bond operator. Again, we find no signatures of ordering from the bond-bond 
structure factor.

Finally, the $\cS-\cI^*$ transition appears to be continuous,
rather than first-order
which would be expected if we had a conventional transition between a 
superfluid and a broken symmetry insulator.
This argument, together with the correlation function results, strongly favor
a non-broken-symmetry ground state for $\cI^*$.  Therefore, we next examine 
the second possibility, that $\cI^*$ is a featureless topologically ordered
Mott insulator (a ``spin-liquid'' in magnetic language).

{\it Topological order in ${\cI}^*$}.---On a lattice with periodic
boundary conditions in both lattice directions,
the subspace of configurations with $n_{\hexagon}=3$ on every kagome 
hexagon, i.e. the ``dimer subspace'' which dominates the boson wavefunction 
for $V \to \infty$,
has topological sectors defined by having, for each lattice direction
$a_{1,2}$, an odd (or even) number of bosons on each row
(``parity sectors''). In this torus geometry and in the restricted Hilbert
space, we thus find four topological sectors labelled by row and column 
parity, such that
any local move which obeys the local constraint $n_{\hexagon}=3$ 
leaves the sector unchanged. In order to change from one sector to 
another one needs to make highly nonlocal boson moves over loops 
which wind around the lattice.

The row/column parities are however not well-defined in
model (\ref{eq:bosonmod}) since at any finite $V$, no matter how large, 
there will be a 
small density of hexagons with $n_{\hexagon} \neq 3$ \cite{footnote.defects}.
We have checked in our numerics, where we start from a configuration in 
the dimer subspace with a certain row/column parity, that the QMC algorithm
generates a small density of particle-hole pair defects in equilibrium,
so that the quantum ground state no longer lies in the ``dimer subspace''. 
However, for a large enough 
system size at a given value of $V/t$ ($L \gtrsim 10$ at $V/t=26$),
we find that our simulations with the longest accessible MC steps do not 
lead to any non-local boson moves which wind around the lattice. Thus the 
winding number 
identically vanishes, and each configuration in the simulation which
lies in the dimer subspace belongs to the same parity sector as the
initial configuration. 
The full ground state accessed by the QMC is, in this sense,
perturbatively connected to the initial parity sector.
This means that four
topological sectors continue to exist, and we can label 
them simply by the row/column parity of that component of the ground 
state wavefunction which lies in the dimer subspace.

For a {\em topologically ordered} insulator, the ground state energy should 
be {\em identical} in each of the four topological sectors (on a torus) 
leading to a ground state degeneracy of four. We have computed the 
energy of the four ground states by starting our simulations from 
configurations in the dimer subspace lying in four different parity sectors.
We find that they are equal within statistical errors, which is strong 
evidence for topological order \cite{footnote.toporder}.

\begin{figure}[t]
\includegraphics[width=3.0in]{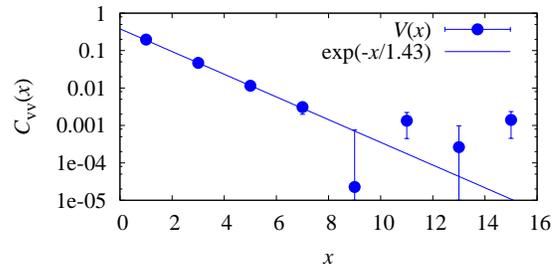}
\caption{ (color online).
Equal-time vison-vison correlation function for $L\!\!=\!\!24$ and
$T\!\!=\!\!t/18$ in
the insulator at $V/t\!\!=\!\!20.5$, showing exponential decay with a
length scale $\xi\!\!=\!\!1.43(5)$.
}
\label{fig:visoncorr}
\vskip -0.2cm
\end{figure}

{\it Spatial vison correlations in ${\cI}^*$}.---A second signature of
$Z_2$ fractionalization is that visons, which are
gapped $Z_2$ vortices in the effective field theory description, should
have exponentially decaying spatial correlations.
The spatial vison-vison correlation function is the expectation
value of a string operator in terms of the spins. For the ring-exchange
model (valid for $V/t \to \infty$), it takes the form \cite{balents1}:
$
  C_{\text {vv}}(r_{ij}) = |\langle 0| 
	   \prod_{k=i}^{j} e^{i\pi n_k}
    |0\rangle|,
\label{eq:vcorr}
$
where $|0\rangle$ denotes the ground state and {\it the product is along some
path on the kagome lattice} that contains an even number of sites, starts at
site $i$, and ends at site $j$, making only ``$\pm 60^\circ$'' turns to
the left or right. $n_k=0,1$ is the number of bosons at site $k$.
The absolute value of the product
is {\em path-independent} in the dimer subspace,
and it is expected to decay exponentially in the topologically 
ordered phase. 
In model (\ref{eq:xxzmod}) at finite $V/t$, ground state no longer lies 
entirely in the dimer subspace, but will mix in configurations with 
particle-hole defects. Correspondingly, the vison operator must include 
correction terms to the above definition since it otherwise becomes 
path-dependent in the presence of such defects.
Both, the corrected ground state wavefunction and the vison 
operator, may be determined order-by-order in a perturbation expansion
\cite{unpub}, undoing the unitary transformation that leads from 
model (\ref{eq:bosonmod}) to a ring-exchange Hamiltonian.

Here evaluate the leading term, correct to ${\cal O}(t/V)$, of
$C_{\text {vv}}(r_{ij})$ by computing the above string expectation value
in the true ground state {\em projected into the dimer
subspace} \cite{unpub}.
As seen from Fig.~\ref{fig:visoncorr}, $C_{\text {vv}}(r_{ij})$
decays exponentially in $\cI^*$, again signaling topological order in
$\cI^*$. At $V/t=20.5$, we estimate a ``decay length'', $\xi=1.43(5)$, which
is comparable to its value at the RK point of the ring-exchange model
\cite{balents1}, $\xi\approx 1$, and to that found by ED \cite{balents2}, 
$\xi\approx 1.7$, in the ring-exchange model with $v_4=0$.

\begin{figure}[t]
\includegraphics[width=3.0in]{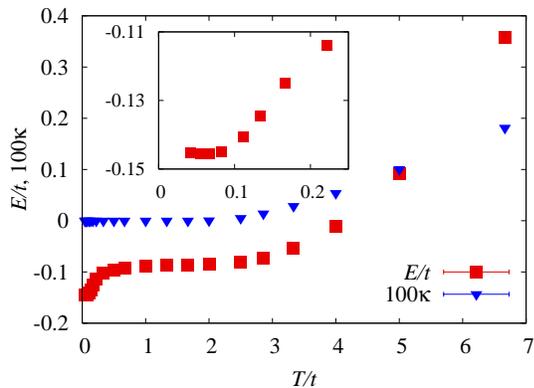}
\caption{ (color online). Energy per site $E$ and compressibility
$\kappa$ versus temperature for $L=24$ ($V/t$=20.5). The energy 
rises exponentially at very low $T$ (see inset) with a wide intermediate 
plateau from $T/t \sim 0.5-3$ (see text). The compressibility is zero 
(within error bar) at low $T$, rising only at $T\sim 3 t$ once gapped
charge (spinon) excitations become relevant.}
\label{fig:energy}
\vskip -0.2cm
\end{figure}

{\it Vison gap}.---To provide further evidence for gapped vison excitations,
we display the temperature dependence of the system energy per site and
compressibility $\kappa = \frac{\beta}{N}
\left\langle \left( \sum_i n_i \right)^2 \right\rangle$
in Fig.~\ref{fig:energy}. The energy exhibits a two-step decrease 
upon lowering temperature, with a distinct intermediate plateau, before
vanishing exponentially at a very low temperature. We identify the first
drop in energy with a freezing out of charge fluctuations below a charge
gap scale, which we confirm by the sharp decrease in $\kappa$ at
this temperature (also shown in Fig.~\ref{fig:energy}). The plateau then
corresponds to a regime where the system dominantly explores
configurations with $n_{\hexagon}=3$ on each hexagon.
(The large difference between the total energy
of the ground state and that in the plateau regime can be explained by the 
macroscopic entropy density of multiple vison excitations \cite{unpub}.) 
Finally at the lowest temperature, the system begins evolving into the 
spin-liquid ground state. Heating up from $T=0$, we therefore identify the 
lowest energy excitations out of the ground state as vison-pair
excitations (since the charge gap is much larger). The temperature dependence 
of the energy then leads to a rough estimate of the single vison gap; for
$V/t=20.5$, we find $E_v/t \sim 0.35(15)$.

{\it Conclusions}.---We have studied the $T=0$
phase diagram of a hard-core boson model with short-range repulsion on the 
kagome lattice using QMC. We find a continuous QPT from a superfluid 
phase to a featureless Mott insulator with increasing 
repulsion. In magnetic language, this corresponds to a ferromagnet-paramagnet
QPT with the paramagnet being a correlated quantum spin-liquid.
This spin-liquid supports gapped visons and exhibits topological order 
characteristic of a $Z_2$ fractionalized phase. The apparently 3D XY
exponents at the $\cI^*-\cS$ transition, $z=1$ and $\nu=0.67(5)$, are
consistent with the QPT arising from condensation of fractionalized
charge-1/2 bosons \cite{senthil:motrunich}. We have 
estimated the vison gap in the spin liquid from the temperature dependence 
of the energy. Finally, the spin liquid
phase is fully gapped and thus stable to weak perturbations 
away from the special choice of exchange couplings in our model.
Although our model is not of direct relevance to known Kagome magnets,
our work represents significant progress
towards understanding realistic Hamiltonians by showing a spin liquid phase in
a model with only local interactions and no special conserved quantities.


We acknowledge support from NSERC (SVI, YBK, AP), CRC, CIAR and
KRF-2005-070-C00044 (SVI, YBK), and an Alfred. P. Sloan Foundation
Fellowship (AP). We thank L. Balents and D.-N. Sheng for useful discussions.


\begin{thebibliography}{99}
\bibitem{footnote.defects}
Such defects correspond to particle-hole pairs 
in real space, which are necessarily bound to each other since the system is 
insulating and charges cannot propagate freely in the ground state.
\end{thebibliography}
\end{document}